\documentclass[sigconf]{acmart}
\AtBeginDocument{%
  }

\usepackage{subcaption}
\usepackage{wrapfig}

\copyrightyear{2023}
\acmYear{2023}
\setcopyright{rightsretained}
\acmConference[RecSys '23]{Seventeenth ACM Conference on Recommender Systems}{September 18--22, 2023}{Singapore, Singapore}
\acmBooktitle{Seventeenth ACM Conference on Recommender Systems (RecSys '23), September 18--22, 2023, Singapore, Singapore}\acmDOI{10.1145/3604915.3610244}
\acmISBN{979-8-4007-0241-9/23/09}

\begin{document}

\title{Learning from Negative User Feedback and Measuring Responsiveness for Sequential Recommenders}

\author{Yueqi Wang}
\email{yueqiw@google.com}
\affiliation{%
  \institution{Google}
  \country{USA}
}
\author{Yoni Halpern}
\email{yhalpern@google.com}
\affiliation{%
  \institution{Google}
  \country{USA}
}
\author{Shuo Chang}
\email{changshuo@google.com}
\affiliation{%
  \institution{Google}
  \country{USA}
}
\author{Jingchen Feng}
\email{jingchenfeng@google.com}
\affiliation{%
  \institution{Google}
  \country{USA}
}
\author{Elaine Ya Le}
\email{elainele@google.com}
\affiliation{%
  \institution{Google}
  \country{USA}
}
\author{Longfei Li}
\email{codyli@google.com}
\affiliation{%
  \institution{Google}
  \country{USA}
}
\author{Xujian Liang}
\email{xujianliang@google.com}
\affiliation{%
  \institution{Google}
  \country{USA}
}
\author{Min-Cheng Huang}
\email{minchengh@google.com}
\affiliation{%
  \institution{Google}
  \country{USA}
}
\author{Shane Li}
\email{lishane@google.com}
\affiliation{%
  \institution{Google}
  \country{USA}
}
\author{Alex Beutel}
\email{alexb@openai.com}
\affiliation{%
  \institution{OpenAI}
  \country{USA}
}
\authornote{work done while at Google}
\author{Yaping Zhang}
\email{yapingzhang@google.com}
\affiliation{%
  \institution{Google}
  \country{USA}
}
\author{Shuchao Bi}
\email{shuchaobi@google.com}
\affiliation{%
  \institution{Google}
  \country{USA}
}

\renewcommand{\shortauthors}{Wang et al.}

\begin{abstract}
Sequential recommenders have been widely used in industry due to their strength in modeling user preferences.
While these models excel at learning a user's positive interests, less attention has been paid to learning from negative user feedback. 
Negative user feedback is an important lever of user control, and comes with an expectation that recommenders should respond quickly and reduce similar recommendations to the user. 
However, negative feedback signals are often ignored in the training objective of sequential retrieval models, which primarily aim at predicting positive user interactions. 
In this work, we incorporate explicit and implicit negative user feedback into the training objective of sequential recommenders in the retrieval stage using a "not-to-recommend" loss function that optimizes for the log-likelihood of not recommending items with negative feedback. 
We demonstrate the effectiveness of this approach using live experiments on a large-scale industrial recommender system. 
Furthermore, we address a challenge in measuring recommender responsiveness to negative feedback by developing a counterfactual simulation framework to compare recommender responses between different user actions, showing improved responsiveness from the modeling change.
\end{abstract}

\ccsdesc[500]{Information systems~Retrieval models and ranking}
\ccsdesc[500]{Information systems~Recommender systems}
\keywords{recommender systems, negative user feedback, loss function, recommender responsiveness, counterfactual simulation} %

\maketitle

\section{Introduction}

In online recommendation platforms, users provide positive and negative feedback on the recommended items through implicit behavioral signals (e.g. clicks, dwell time) and explicit actions (e.g. likes, dislikes, dismissals). 
Negative user feedback, in particular, 
often comes with a strong expectation that the recommender will respond quickly and 
reduce similar recommendations to the user \cite{ricks2022does}. 
Thus, it is important for a responsible recommender system to learn from negative feedback and adjust personalized recommendations based on these signals. 
This is especially true for real-world applications with dynamic user preferences and ever-increasing item spaces, such as short-form content platforms where recommenders crafts the personalized feed, and users consume large numbers of items per session.

Leveraging negative feedback in recommenders is an active area of research ~\cite[e.g.,][]{frolov2016fifty,seo2022siren,hamed2012impact,paudel2018loss,park2022exploiting,zhao2018recommendations,cena2023deal}. In this work, we focus on modeling negative user feedback in sequential recommenders, a type of model that learns
sequential user interaction patterns 
to generate recommendations
and is widely used in industry \cite{wang2019sequential,zhang2019deep,chen2019top,tang2019towards,donkers2017sequential,hidasi2015session}. Sequential recommenders
in the retrieval stage have largely focused on the objective of predicting positive interactions, with less attention on learning from negative feedback. 
Particularly, negative user feedback is often not used in the training objective to teach the model what not to recommend for a given user state, 
leaving the handling of such signals to downstream components (e.g. ranking stage).
Although negative interactions may appear in input features, 
softmax negative samples \cite{yang2020mixed, chen2023revisiting}, or auxiliary tasks \cite{chen2021user} for sequential retrieval models, 
the lack of a dedicated training objective makes it hard to effectively adjust model predictions when the recommended items do not align with a user's interest.

Furthermore, measuring a recommender's responsiveness to negative user feedback is challenging. 
In observational data, it is difficult to disentangle user feedback from other correlated signals and quantify its influence on subsequent recommendations due to the complex user-recommender feedback loop.
While online user metrics are the north-star, 
knowing how the system behaves upon negative feedback is essential for a responsible and interpretable recommender. 

In this work, we incorporate negative user feedback into the training objective of a large-scale industrial sequential recommender, and develop a counterfactual simulation framework to measure the responsiveness to negative feedback: 

\begin{itemize}
    \item We introduce a "not-to-recommend" loss function in sequential recommenders where items with negative user feedback are used as  training labels to optimize for the log likelihood of not recommending unwanted items. %
    This adds an additional term to the training objective to jointly learn from both positive and negative user feedback.
    \item Using live experiments on an industrial recommender system serving billions of users, we show improved user experience from incorporating explicit and implicit negative feedback into the recommender's training objective.
    \item We present a counterfactual simulation method that compares recommender responses between hypothetical user actions 
    to measure the reduction of similar recommendations upon negative feedback. This shows the proposed training objective improves the recommender's responsiveness to negative user feedback.
\end{itemize}

\section{Training objective for negative user feedback}
Below we describe our approach for incorporating negative user feedback into the training objective of sequential recommenders, specifically for the retrieval task of predicting the next set of items to recommend from a large corpus.

\begin{figure}
    \includegraphics[width=0.44\textwidth]{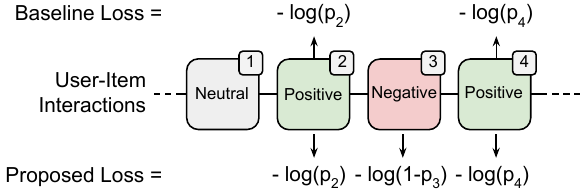}
    \caption{Incorporating negative user feedback into the training objective of sequential recommenders. The diagram shows a sequence of user-item interactions. Items with positive or negative feedback are used as training labels.}
    \Description{Incorporating negative user feedback in the training objective.}
    \label{fig:negative_label_diagram}
\end{figure}

\textbf{Loss Function.}
Sequential retrieval models are typically trained for the task of predicting positive interactions using training objectives such as the cross-entropy loss or REINFORCE policy gradient \cite{beutel2018latent,chen2019top}. 
Here, we start with a baseline model that uses the cross-entropy loss with positive user feedback as training labels. 
Given a training example $i$ containing a user $u_i$'s positive interaction with a label item $y_i$, the loss function is the negative log likelihood $\mathcal{L}_i = -log(p(y_i|\mathbf{s}_i))$, where $\mathbf{s}_i$ represents the user state vector computed by a neural network that summarizes the user's interaction history. 
This loss function optimizes the conditional probability $p(y_i|\mathbf{s}_i)$ of recommending the label item $y_i$ out of a large corpus given the user state $\mathbf{s}_i$.
The probability term is expressed as a softmax: 
$p(y_i|\mathbf{s}_i)= \exp{(\mathbf{s}_i^\intercal \mathbf{v}_{y_i})} /\sum_{y_i' \in \mathcal{A}} \exp{(\mathbf{s}_i^\intercal \mathbf{v}_{y_i'})}$
, where $\mathcal{A}$ is the item space and $\mathbf{v}_{y_i}$ represents the item embedding vector.
In practice, sampled softmax is used in training and nearest neighbor search in serving to handle extremely large item spaces \cite{chen2019top}.

To learn from negative user feedback, we introduce a "not-to-recommend" loss function which is the negative log likelihood of not recommending an item, i.e. 
$\mathcal{L}_i = -log(1-p(y_i|\mathbf{s}_i))$. 
This objective allows the model to directly leverage negative user feedback as negative training labels,
and works alongside with the existing loss function for positive labels.
Adding a weight $r_i$ for each positive label and $w_i$ for each negative label, we get the overall loss function:
\begin{equation}
\mathcal{L} = - \sum_{i \in D_{\text{pos}}} \biggl( r_i \cdot \log(p(y_i | \mathbf{s}_i)) \biggr) - \sum_{i \in D_{\text{neg}}} \biggl( w_i \cdot \log(1 - p(y_i | \mathbf{s}_i)) \biggr),
\end{equation}

where $D_{\text{pos}}$ and $D_{\text{neg}}$ are the positive and negative training labels, respectively (Fig \ref{fig:negative_label_diagram}). Minimizing this loss corresponds to maximizing the joint probability of recommending items with positive feedback and not recommending those with negative feedback.
This enforces stronger alignment with user interests compared to using positive labels alone.

\begin{figure*}[t!]
    \centering
    \begin{subfigure}[t]{0.4\textwidth}
        \centering
        \includegraphics[width=0.75\textwidth]{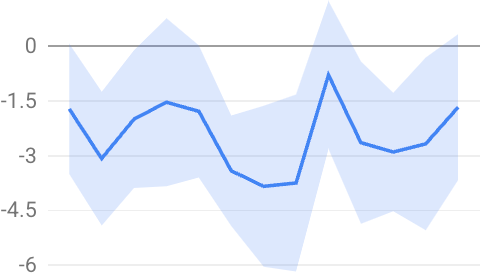}
        \caption{Dislike rate (homepage)} 
        \Description{Dislike rate (homepage)}
        \label{fig:dislike_rate_home}
    \end{subfigure}
    \begin{subfigure}[t]{0.4\textwidth}
        \centering
        \includegraphics[width=0.75\textwidth]{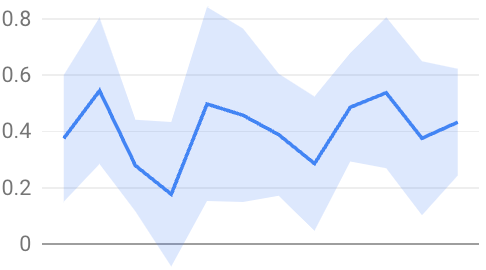}
        \caption{Overall enjoyment (immersive feed)}
        \Description{Overall enjoyment (immersive feed)}
        \label{fig:overall_enjoyment_ip}
    \end{subfigure}
    \vspace{-3mm}
    \caption{Live experiment results for modeling dislike (a) and skip (b) feedback. X-axis is date and y-axis is the relative metric change in percentage. The light blue area represents the 95\% confidence interval.}
    \Description{Experiment results for modeling dislike (a) and skip (c) feedback. X-axis is date and y-axis is the relative metric change in percentage. The light blue area represents the 95\% confidence interval.}
    \label{fig:live_experiment}
\end{figure*}

The "not-to-recommend" loss function addresses several practical issues of modeling negative user feedback in the training objective. For example, using cross-entropy loss $\mathcal{L}_i = - log(p(y_i|\mathbf{s}_i))$ with negative-valued label weights could also reduce the probability of recommending unwanted items, but this leads to gradient blow-up when $p(y_i|\mathbf{s}_i) \rightarrow 0$. In principle, reinforcement learning objectives support assigning negative reward values to negative training labels. In fact, we could replace the loss term for positive labels by a REINFORCE objective \cite{chen2019top}. However, for negative labels, using negative rewards in REINFORCE recommenders faces a practical challenge, where gradient blow-up may still occur when $p(y_i|\mathbf{s}_i) \rightarrow 0$ even after off-policy correction due to the extremely large item spaces in industry settings. The "not-to-recommend" loss function circumvents these issues as the gradient is guaranteed to be finite when $p(y_i|\mathbf{s}_i) \rightarrow 0$.
An alternative approach is to include and upweight negative feedback items among the softmax negative samples of adjacent positive labels. Compared to this approach, the proposed method decouples the loss terms for positive and negative user feedback, and the gradient allows more targeted learning from negative labels.

\textbf{Negative Training Labels and Input Features.}
Both explicit and implicit negative user feedback can serve as negative training labels.
The label weights can be tuned according to the feedback strength, signal density, and the relative loss value magnitude compared to positive labels.
On our platform,
we consider explicit dislike actions and implicit skip signals %
as negative training labels.
To represent a user's past negative interaction in model input,
each item in the model's input sequence encodes dislikes by a binary feature, and skips as part of the dwell time feature.

\section{LIVE EXPERIMENTS}

We incorporate the "not-to-recommend" loss into a recurrent neural network-based sequential recommender \cite{beutel2018latent, chen2019top} of a short-form content platform serving billions of users, and conduct live experiments 
to evaluate the real-world impact of modeling negative user feedback. 
In our system, the sequential recommender is part of the retrieval stage and recommends hundreds of items from a huge corpus, followed by a ranking stage before the best items are shown to the user \cite{covington2016deep}. 
In practice, negative user feedback is also handled by downstream components, such as ranking models and heuristics 
\cite{zhao2019recommending, mosseri2023instagram, twitter2023algo}. 
Even so, we see meaningful end-to-end results from this model change. 
We show that learning from negative user feedback in retrieval can improve recommendation by (1) reducing the chance of unwanted recommendations slipping through the system, and (2) enhancing the model's alignment with user interests to generate better recommendations.
The metric changes reported below are significant under a 95\% confidence interval.

\textbf{Modeling Explicit Negative Feedback.} 
In this experiment, we incorporate explicit dislike feedback into the model's training objective. 
On the product's homepage, the experiment model using dislikes as both input feature and training labels reduces dislike rate by 2.44\% compared to a baseline of not using dislike signals (Fig \ref{fig:dislike_rate_home}). 
This effect is much larger than only using dislike as input feature but not training labels (-0.34\%, not significant), or 
a heuristic solution that excludes disliked items from the model's input sequence (-0.84\%).
Repeated dislike rate on the same creator decreases by 9.60\%, suggesting that the model reduces similar recommendations after negative feedback. The experiment model reduces dismissing users by 2.05\% whereas viewers only drop by 0.14\%. 
Dismissals are not explicitly modeled in our treatment, and this change indicates improved user experience. 
In the immersive feed of our product, the experiment model reduces dislike rate by 1.10\%
and repeated dislikes on the same creator by 7.10\%. 
Using model-based satisfaction metrics \cite{christakopoulou2022reward}, satisfied consumption stays the same (+0.00\%) and unsatisfied consumption decreases by 0.35\%.

\textbf{Modeling Implicit Negative Feedback.}
The implicit skip signal has a much higher density than explicit dislikes. 
We hypothesize that a higher coverage of negative user interests can improve 
the overall recommendation quality.
In this experiment, we include skipped items as negative training labels on the immersive feed surface.
We observe a 0.40\% increase of overall user enjoyment (Fig \ref{fig:overall_enjoyment_ip}), 
and a 0.61\% increase of daily active users with at least 1 hour of daily activities. 
In addition, we see a 1.44\% reduction of repeated skips from the same creator and a 0.88\% increase of diversity. 

Together, the results show that incorporating explicit negative feedback in the training objective reduces negative user experiences, and modeling implicit negative feedback improves overall user enjoyment.

\begin{figure*}%
    \centering
    \begin{subfigure}[b]{0.58\textwidth}
        \centering
        \includegraphics[width=\textwidth]{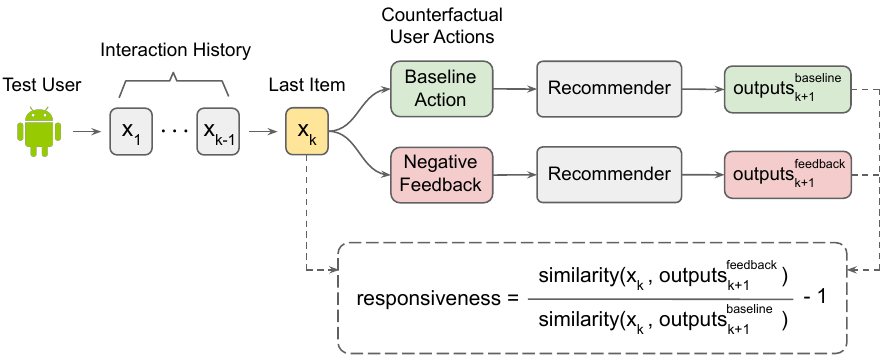}
        \caption{Counterfactual simulation for measuring responsiveness to user feedback.} 
        \Description{Counterfactual simulation framework for measuring a recommender's responsiveness to user feedback.}
        \label{fig:simulation_diagram}
    \end{subfigure}
    \hfill
    \begin{subfigure}[b]{0.39\textwidth}
        \centering
        \includegraphics[width=\textwidth]{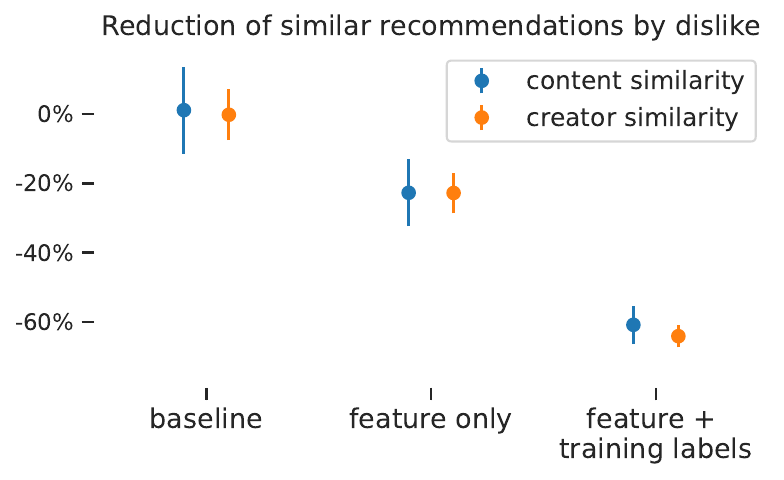}
        \caption{Improved responsiveness to dislikes.}
        \Description{Improved responsiveness to dislike feedback.}
        \label{fig:simulation_result}
    \end{subfigure}
    \vspace{-3mm}
    \caption{Measuring recommender responsiveness by counterfactual simulation.}
    \Description{Measuring recommender responsiveness by counterfactual simulation.}
    \label{fig:simulation_diagram_and_results}
\end{figure*}

\section{MEASURING RESPONSIVENESS BY COUNTERFACTUAL SIMULATION}
The live experiment shows 
the overall benefits of modeling negative feedback,
but does not provide a direct measurement of how responsive the recommender is to such feedback signals.
We aim to measure the responsiveness as how much the recommender reduces similar recommendations when a user reacts negatively to an item. However, this recommender behavior cannot be assessed by offline model accuracy, online user metrics, or correlative log data analysis due to the complex user-recommender interactions. 
To directly measure recommender responsiveness, we need to factor out the confounding user behaviors, 
and compare the causal effects of counterfactual user actions on recommender responses.

Here, we develop a framework that uses simulated users \cite{yao2021measuring} to measure responsiveness to user feedback (Fig \ref{fig:simulation_diagram}). 
Each user follows a random behavior trajectory to consume a series of $k-1$ recommended items. 
Upon the $k$th item, we simulate multiple counterfactual actions (e.g. reacting negatively to it, or not doing so) under identical interaction history.
In each counterfactual branch, we observe the recommender outputs for the $(k+1)$th step and compute their similarity with the last interacted item. 
Using independent measures of content-based and creator-based similarity,
we define the similarity score as the proportion of recommendations from the same content cluster or creator as the last interacted item. 
The recommender responsiveness is then computed as the relative change of this similarity score between the counterfactual actions of providing vs. not providing negative feedback. This tells us how much the recommender reduces similar recommendations upon negative feedback, and can be used to evaluate modeling changes.

We apply this approach to measure the recommender's responsiveness to the dislike action, which has been difficult to assess due to its sparseness. 
For counterfactual actions, we consider a positive interaction baseline (long dwell time) vs. adding a dislike action on top of it. We run 2000 simulations with $k=50$.
The baseline model without using dislike signals has no responsiveness, and would rely on downstream system components to respond to dislikes. 
Adding dislike input feature alone 
leads to a significant but limited reduction of
similar recommendations after dislikes (-22.7\%/-22.8\% by content/creator similarity), indicating that the dislike action represents an intrinsic user dissatisfaction. 
When using dislikes in both training labels and input features, the model yields a larger reduction of similar recommendation upon dislikes (-60.8\%/-64.1\% by content/creator) (Fig \ref{fig:simulation_result}). 
These results show that incorporating explicit negative user feedback in the training objective 
improves the model's responsiveness to such feedback.

\section{CONCLUSIONS}
We introduce a generic training objective to learn from explicit and implicit negative user feedback in sequential recommenders, and a simulation framework for measuring the recommender's responsiveness to negative feedback. 
The results show effective reduction of unwanted recommendation and improvements of user experience. The proposed modeling approach and responsiveness measure can be generalized to other feedback signals and recommender models.

\section{Presenter Bio}
Yueqi Wang is a software engineer at Google working on recommender systems.

\noindent
Yoni Halpern is a software engineer at Google Research working on Responsible AI.

\bibliographystyle{ACM-Reference-Format}
\bibliography{recsys2023_arxiv}

\end{document}